\documentclass[12pt]{iopart}
\usepackage{graphicx}
\usepackage{xcolor}
\usepackage{ulem}
\begin{document}

\title[R.\ C.\ dos Santos \textit{et al.}]{Influence of 
pinning centers of different natures on surrounding vortices}

\author{Rodolfo Carvalho dos Santos$^{1}$, Elwis Carlos Sartorelli Duarte$^{1}$, 
Danilo Okimoto$^{1}$, Alice Presotto$^{1}$, Edson Sardella$^{2}$, 
Maycon Motta$^{3}$, Rafael Zadorosny$^{1}$}

\address{$^{1}$Departamento de F\'{\i}sica e Qu\'{\i}mica, Universidade 
Estadual Paulista (UNESP), Faculdade de Engenharia, Caixa Postal 31, 
15385-000, Ilha Solteira, SP, Brazil}

\address{$^{2}$Departamento de F\'{\i}sica, Universidade Estadual 
Paulista (UNESP), Faculdade de Ci\^ encias de Bauru, 
Caixa Postal 473, 17033-360, Bauru, SP, Brazil}

\address{$^{3}$Departamento de F\'{\i}sica, Universidade Federal 
de S\~ao Carlos (UFSCar), S\~ao Carlos, SP, Brazil} 

\ead{rafael.zadorosny@unesp.br}
\vspace{10pt}
\begin{indented}
\item[]May 2021
\end{indented}

\begin{abstract}
Studies involving vortex dynamics and their interaction with 
pinning centers are an important ingredient to reach higher 
critical currents in superconducting materials. The vortex 
distribution around arrays of engineered defects, such 
as blind and through holes, may help to improve the 
superconducting properties. Thus, in this work, we used 
the time-dependent Ginzburg-Landau theory to investigate 
the vortex dynamics in superconductors of mesoscopic 
dimensions with a large central square defect with three 
different configurations: (i) a hole which passes through 
the sample (interface with the vacuum); (ii) a  
superconducting region with lower critical temperature ($T_c$); 
and (iii) a region with a more robust 
superconductivity, i.e., with a higher $T_c$. 
Such systems can be envisaged as elementary building blocks 
of a macroscopic decorated specimen. Therefore, 
we evaluated the influence of  
different interfaces on the vortex dynamics and their 
effects in the field-dependent magnetization and 
time-dependent induced electric potential variation. 
The results show that the lower critical field is 
independent from the nature of the defect. However, 
the currents crowd at the vertices of the through 
hole producing a lower degradation of the local 
superconductivity, which may increase the upper 
critical field. On the other hand, the last type 
of defect can be used to control the vortex dynamics 
in the main superconducting region around the defect with more accuracy.
Whereas the first two defects are attractive for 
the vortices, the third type is repulsive for them, 
being needed several vortices penetrated in the 
superconducting matrix to have vortices penetrated into it.

\end{abstract}

%
%
%
\submitto{\SUST}
%
%

\section{Introduction}\label{sec:sec1}

In type-II superconductors, the viscous motion of 
vortices generates a resistive state producing heat 
and might lead the material to the normal state. Then, 
vortices must be kept static to avoid the premature 
destruction of the superconductivity. One 
way to trap such fluxoids is the 
introduction of attractive potentials into the 
superconducting sea, the so-called pinning centers 
(PCs), which can maintain the vortices pinned against 
the Lorentz Force produced by both shielding and transport 
currents circulating around the material~\cite{blatter1994rmp,CampbellEvetts1972}. 
On one hand, PCs consist of defects naturally 
present, which occurs during the material preparation, 
such as twin-boundaries~\cite{kwok1990direct}, oxygen 
vacancies~\cite{vargas1992flux}, and grain boundaries~\cite{pande1976model}. 
On the other hand, a variety of works have been 
reported the use of artificial PCs, randomly or 
regularly ordered, as the introduction of arrays of 
magnetic dots~\cite{martin1997flux,martin1999artificially} 
and through or blind holes~\cite{moshchalkov1998pinning, weiss1994magnetotransport}. 
In several cases, lithographic techniques based on electron, 
ion, and neutron beam radiation have been used to produce 
both through holes, also know as antidots (ADs)~\cite{A1,A2,A3}, 
as well as blind holes (BHs)~\cite{A4,A5,A6,A7,A8}.

As a matter of fact, samples patterned with BHs 
present a pinning force weaker in comparison to specimens 
decorated with ADs~\cite{A4,raedts2004flux-pinning,raedts2004flux}. 
Additionally, both vortex states and their interaction 
with the interfaces between the defect and the 
superconductor can be accessed. In Ref.~\cite{A5}, a 
Nb thin film was irradiated by reactive ion etching to 
form a regular array of circular BHs with radius in the 
range $0.15~\mu{m}<R<2.2~\mu{m}$. The authors showed that 
the trapped vortices form a shell-like structure following 
the geometry of the defect. Moreover, the number of vortices 
that nucleate inside the hole is not sensitive to the thickness 
of the BHs. On the other hand, using the Ginzburg-Landau 
formalism~\cite{A4}, it was shown that different vortex 
states as dimer, trimer, and other shell-like structures, 
are available into the BHs depending on the number 
of trapped vortices. For large BHs (${R} = 6\xi(0)$, where $\xi(0)$ 
is the coherence length at zero temperature), the average 
distance between vortices depends on both the radius of 
that structure and the total number of vortices penetrated 
in the sample. Besides that, for very large BHs (${R} = 30\xi(0)$), 
vortices are arranged in an Abrikosov lattice at the center 
of the BH for vorticities counting hundreds of flux quanta. 
However, in the BH contour, the vortices form a shell-like 
distribution. The above-cited works showed that the vortices 
into the defects behave independently of those in the main 
superconducting matrix.\footnote{By \textit{superconducting 
matrix}, we mean the main region around the defect which is 
either a superconductor of lower or higher critical 
temperature ($T_c$).} Furthermore, it is interesting to 
note that inside the BHs, the vortices experience c
onfinement effects similar to those presented by 
mesoscopic systems~\cite{palacios1998vortex, palacios1998vorteXXX, geurts2010vortex, pereira2011vortex}.

Concerning heterostructures with thickness modulation, 
Brisbois~\textit{et al.}~\cite{brisbois2017prb} have 
fabricated by e-beam lithography two 140-nm thick Nb films 
with a central macroscopic defect shaped like a polygonal 
thickness step, one excavated with a thinner thickness (80 nm) 
and another one with a thicker central region (200 nm). 
In the former case, vortices penetrate more profound due to the 
reduction of the thickness-dependent critical current in 
its thinner center, whereas in the latter, the thicker defect 
behaves as a barrier for the incoming fluxoids. Using the 
time-dependent Ginzburg-Landau (TDGL) formalism, the authors 
traced the vortex trajectories and confirmed these features. 
Moreover, these behaviors were also observed in abrupt 
flux penetration phenomena, the so-called flux avalanches, 
in which sudden bursts of ultrafast dendritic flux rush 
into the sample. Another way to substantially enhance the 
protection against avalanches consists of the addition of 
a thin Nb layer before the reactive deposition of NbN, 
thus forming a NbN/Nb hybrid~\cite{Pinheiro2020}.

In a recent work~\cite{A9}, the TDGL formalism was applied 
to study the vortex dynamics in thin and large superconducting 
films with a hexagonal array of PCs. The results were 
contrasted with experimental measurements of the critical 
current density of a patterned MoGe thin film. The effects 
of the pinning force were studied for different types of 
defects, their sizes, and displacement on the sample. 
It was shown that $J_{c}(H)$ is defect-dependent, being 
higher for BHs with diameter ${d}=300~\textrm{nm}$, and a 
depth ${h}=300~\textrm{nm}$. Also, different vortex dynamics 
were associated with matching fields, i.e., the 
commensurability effects between the vortex lattice with the 
pinning lattice that enhances the critical current density~\cite{fiory1978mf}.

Vortex pinning and matching fields were also reported in 
superconductors with \textit{antipinning centers}, i.e., 
samples decorated with superconducting dots. In Ref. \cite{navarro2015enhancements}, 
an increase of the
the authors described an increase of the critical current density, 
$J_c$, as a consequence of a vanadium film deposited on an array of 
Nb dots for temperatures below $3~\textrm{K}$. Additionally, 
Gillijns and coworkers\cite{gillijns2007superconducting} 
deposited an array of Pb microrings over an Al film. For 
zero-field cooling measurements, the microrings repel the 
vortices to the interstitial pinning potentials; conversely, 
an attractive interaction between vortices and microrings 
takes place for field cooling processes. Such change of 
antipinning behavior was also reported by Carreira 
\textit{et al.}\cite {carreira2014superconducting} in Nb films 
with a triangular array of vanadium dots. Besides 
that, the equilibrium vortex configurations were 
described by Berdiyorov and coworkers~\cite{berdiyorov2008pillars} 
in the framework of Ginzburg-Landau theory. Pillars 
were considered as antipinning centers, and some 
theoretical results were experimentally 
demonstrated in Nb film etched with a square array 
of circular pillars. The authors showed that 
the radius and thickness of the pillars 
induces a plethora of vortex structures in the i
nterstitial sites. Also, for larger pillars, vortices 
penetrate them forming shell structures by inducing 
the same symmetry to the interstitial ones.

In general, it is noted that studies of the vortex interaction 
with defects are an issue of great importance from fundamental 
to applied physics of the superconducting materials. Besides that, 
most used systems are in bulk or film forms, and confinement 
effects are present only inside the BH-like defects or antipinnings.
Thus, in the present work, we simulated, using the TDGL formalism, 
square samples decorated with different square defects: an AD, a blind 
hole, and a strong dot. The first one is described as a superconducting 
matrix with a through hole in its center with superconducting-vacuum 
interfaces. The second and the third ones are composed of central 
superconducting regions with a lower critical temperature ($T_c$) 
and a higher $T_c$, respectively. Hence we focus our investigation 
on the influence of these different types of mesoscopic defects 
in the vortex matter of superconducting samples at the nanoscale.

\section{Theoretical Formalism}\label{sec:sec2}

In the TDGL formalism~\cite{Schmid}, the complex superconducting 
order parameter, $\psi$, and the magnetic vector potential, 
\textbf{A}, are described as a function of time. Within such 
formalism, an electric potential $\Phi$ can be taken into account. 
Thus, it is possible to study non-equilibrium states of 
the vortex matter. Those equations in dimensionless units are given by,

\begin{equation}
u\bigg(\frac{\partial}{\partial t} +i\Phi\bigg)\psi = (-i\nabla- \textbf{A})^2 \psi + (g(\textbf{r}) - T - \vert\psi\vert^2 )\psi,
\end{equation}

\begin{equation}
\bigg(\frac{\partial \textbf{A}}{\partial t} + \nabla \Phi\bigg) = \textbf{J} - \kappa^2 \nabla\times\nabla \times \textbf{A},
\end{equation}

\begin{equation}
\textbf{J} = \textrm{Re}(\psi^* (- i\nabla - \textbf{A}) \psi )
\end{equation}
where $\textbf{J}$ is the superconducting current 
density. The function $g(\textbf{r})$  
defines the defect regions. For $g~>~1$, the defect is a 
superconducting material with $T_c$ larger than that of 
the matrix, whereas for $0~\leq~g~<~1$, the local $T_c$ is lower 
than that of the matrix; $\kappa$ is the Ginzburg-Landau 
parameter defining the superconducting material. 
Equations (1), (2), and (3) are dimensionless, the distances 
are written in units of $\xi(0)$; the fields in units of the upper 
critical field at zero temperature, $H_{c2}(0)$; $\textbf{A}$ 
is in units of $H_{c2}(0)\xi(0)$, and time in units of 
the characteristic GL time, 
$t_{GL}(0)~=~\frac{\pi\hbar}{8k_{B}T_{c}{u}}$. Thus, ${u}$ is 
the ratio between $\frac{\tau_{|\psi|}}{t_{GL}}$, 
where $\tau_{|\psi|}$ is the relaxation time of $|\psi|$. 
Following the microscopic derivation of the TDGL equation 
presented by L. P. Gor'kov and G. M. Eliashberg~\cite{GorkovEliashberg}, ${u} = 12$ 
for a superconductor with large concentration of 
paramagnetic impurities. However, in situations where the studied 
phenomena are not sensible to the relaxation process, 
it can be set ${u} = 1$ (or less) to minimize computational 
time~\cite{Vodolazov2000,Vodolazov2003}. 

Considering a better analysis involving the interaction 
between vortex dynamics and mesoscopic defects, in 
this work, the time-dependence of the magnetization $M(t)$, 
induced voltage $V(t)$, and free energy $E(t)$, at non-equilibrium 
processes were studied. In normalized units, such 
equations are given as follows.

\begin{equation}
{V} = -\frac{\partial}{\partial{t}} \int_{C} \textbf{A}\cdot{d}\textbf{l},
\end{equation}

\begin{equation}
\textbf{M} = \textbf{B} - \textbf{H},
\end{equation}

\begin{equation}
{E} =  \bigg(\frac{1}{2}|\psi|^{2}-(g(\textbf{r})-{T})\bigg)|\psi|^{2} +
\bigg|(-i\nabla- \textbf{A})\psi\bigg|^{2} +  \kappa^{2}(B - {H})^{2}.
\end{equation}
Here, $H$ is the external magnetic field. Since the studies 
were carried out with no transport current, we applied the 
Coulomb gauge $\Phi = 0$ for all times and positions. All equations were 
discretized following the link variable method~\cite{A11}, 
an important tool that preserves the gauge invariance 
of the TDGL equations.

\subsection{Sample details}\label{subsec:subsec2.1}

\begin{figure}[!h]
    \centering
    \includegraphics[width=0.75\textwidth]{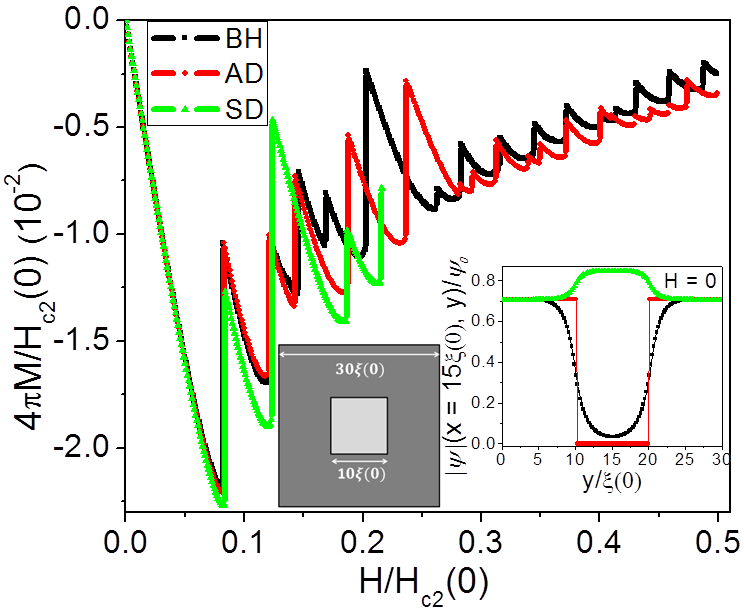}
    \caption{Magnetization as function of the applied magnetic 
    field of the simulated samples. The left inset shows an 
    schematic view of the systems, and the right inset shows 
    the distribution of $|\psi|$ in the middle of the samples, 
    i.e., along the $y$-axis at $x= 15\xi (0)$.}
    \label{fig1}
\end{figure}

We simulated three superconducting square samples with lateral 
size $L = 30 \xi(0)$ with a concentric square defect of 
dimension $l=10 \xi(0)$ (see the left inset of Figure~\ref{fig1}). The 
discretization of equations (1 - 6) was made in a uniform mesh 
taking into account with five points per coherence length, i.e., 
$\Delta x = \Delta y = 0.2 \xi (0)$. Three different types 
of defects were considered: (i) a blind hole, which consists of 
a material with $T_c$ lower than that one of the superconducting 
matrix; (ii) a through hole; and (iii) a material whose $T_c$ 
is greater than that of the superconducting matrix, which hereafter 
we refer to as a \textit{strong defect} (SD). It was also considered 
$\kappa = 5$ (which refers to a Pb-In alloy)~\cite{poole1995superconductivity}, 
and the temperature was set at $0.5 T_c$. The external 
magnetic field was increased in steps of 
$\Delta H = 10^{-3} H_{c2}(0)$ from $H=0$ to $H=0.5 H_{c2}(0)$. 
Four our purposes, it was sufficient to go up to $H = 0.216 H_{c2}(0)$.

Hereafter, the specimens will be referred by their 
associated defects, i.e., the BH, AD, and SD, for 
those with a blind hole, an antidot, and a strong defect, 
respectively. The effect of the nature of the defects over $\psi$ 
is shown in the right inset of Figure~\ref{fig1}. The distribution 
of the superconducting order parameter was taken at $y$-axis, and $x=15\xi(0)$.  

\section{Results and discussion}\label{sec:sec3}

\subsection{Equilibrium aspects - Vortex configurations}
\label{sec:subsec3.1}

\begin{figure}[!h]
    \centering
    \includegraphics[width=0.5\textwidth]{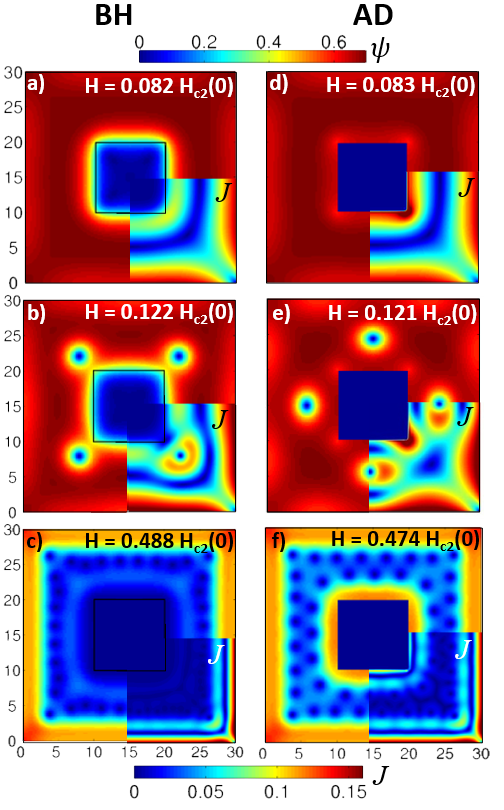}
    \caption{Comparison of the order parameter $|\psi|$ between 
    BH and AD samples for the equilibrium state of three values of 
    the external applied field. Each of the panels also 
    shows the distribution of the  
    absolute value of the current density $J$ 
    in the samples.}
    \label{fig2}
\end{figure}

Figure~\ref{fig1} shows the behavior of the magnetization 
as a function of the applied field for the three simulated samples, 
each one with a different type of central defect mentioned 
previously. As can be seen, no matter what is the type of the defect, 
the lower critical field ($H_{c1}$) is approximately the same, 
namely, $0.082 H_{c2}(0)$, $0.083 H_{c2}(0)$, and $0.084 H_{c2}(0)$ 
for BH, AD, and SD samples, respectively. The magnetic 
behavior of the BH and AD samples is quite similar up to the 
third vortex penetration. On the other hand, the $M(H)$ curve 
of the SD sample does not match with those ones of the other 
samples for all $H>H_{c1}$. It is important to mention 
that the nature of the defects affects the magnetic 
response of superconducting films. For example, flux 
avalanches~\cite{raedts2004flux, brisbois2017prb, silhanek2004prb}, 
which are usually undesired events, can be influenced 
by using different types of engineered pinning as shown 
in Ref. ~\cite{montero2003}.

Figure~\ref{fig2} shows snapshots of $|\psi|$, and the 
absolute value of the shielding currents, $J$, 
positioned at the fourth quadrant of each image, for 
both the BH and AD samples. The stationary states after 
the first vortex penetration are shown in panels (a) and (d). 
As one can see, a more degraded superconducting region 
takes place around the BH defect than in the vicinity of 
the AD one. In addition, the distribution of 
currents significantly changes from one type of defect to another. 
Moreover, the distribution of vortices inside the defects 
is rather different. While in the AD sample, the flux is 
continuously distributed into the defect. In the other case, they seem to 
be spread inside the BH. Just after the second penetration 
at $H = 0.122 H_{c2}(0)$ for BH sample, and 
$H = 0.121 H_{c2}(0)$ for AD one, 
four more vortices are found at the vertices of 
the defect for the former case, whereas they are 
positioned in the midpoint of the square of the later one 
(see panels (b) and (c)). This can be explained by the fact 
that the current density is much larger for the AD than for the 
BH. Such current crowding effect~\cite{ClemBerggren2011} will 
be discussed in more details latter on. Finally, the last 
analyzed penetration occurred at $H = 0.488 H_{c2}(0)$ 
for BH, and $H = 0.474 H_{c2}(0)$ for AD samples. 
Up to this stage, BH and AD samples accumulate 56 and 54 penetrated vortices, respectively.
Although these numbers are similar, the BH defect 
presents twice as many vortices (20) as the AD one. 
This result is in opposition to what was found 
in Ref.~\cite{raedts2004flux}. However, we must 
emphasize that in this work, the authors considered an infinite 
film with a regular array of defects. This points out that, in 
mesoscopic materials, smooth surfaces between the superconducting 
region and the defect create a lower barrier to the penetration 
of more vortices in a BH than in an AD.

Still concerning Figure~\ref{fig2} another import remark 
must be stressed. From panels (c) and (f) 
we can observe that around the BH, the superconductivity 
is suppressed, and the vortices are disposed of in a shell-like 
structure (panel (c)). On the other hand, for the AD sample 
(panel (f)), the superconductivity is more preserved  
around the defect, and the vortices are 
distributed in two rows forming a near triangular lattice. 

\begin{figure}[!h]
    \centering
    \includegraphics[width=0.5\textwidth]{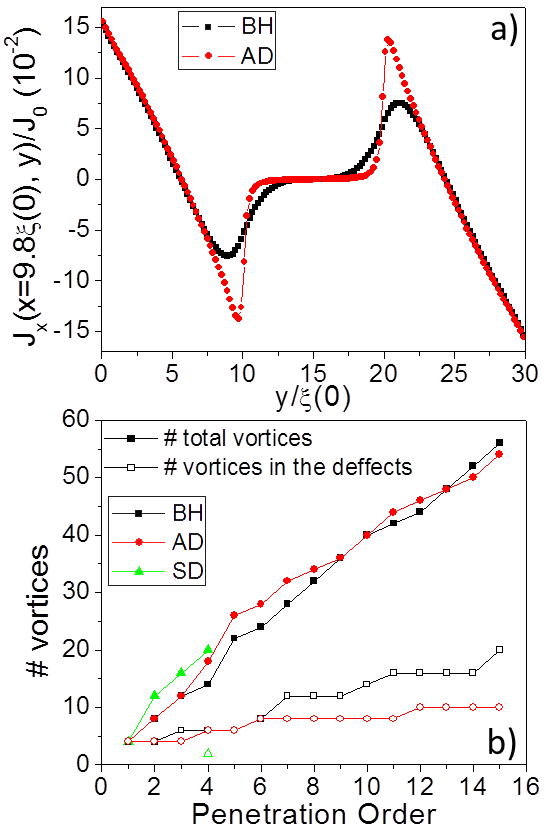}
    \caption{ (a) Variation of the $J_x$ as a function of $y$ 
    for $x=9.8\xi(0)$. The current at the edges of the defects 
    is more significant and sharper for the AD. (b) The closed 
    symbols represent the total vorticity in the samples in 
    each penetration event, and the open symbols show the 
    vorticity of the defects.}
    \label{fig3}
\end{figure}

In panel (a) of Figure~\ref{fig3} is shown the 
$x$-component of the current density of the superconducting current 
along the $y$-axis for the fixed value of $x=9.8 \xi(0)$ 
immediately after the first penetration of vortices
for both BH and AD samples. 
First, notice that the defects' currents circulate opposite 
to the shielding ones at the samples' borders, as expected. 
As can be seen, the intensity of $J_x$ is more significant 
for the AD defect by showing a sharper distribution near the 
edge of the defect. Such behavior implies in a stronger 
repulsive interaction with 
external vortices than that one presented by the BH. 
In panel (b) of Figure~\ref{fig3}, we exhibit the 
total number of vortices in the samples 
after each vortex penetration. All systems, 
the mixed state sets in with vorticity four (4 vortices), which 
means that this first event is only due the 
the geometry of the sample. The subsequent penetrations 
show that both the geometry and the characteristic of the  
defects are essential to control the configuration of 
vortices.

\begin{figure}[!h]
    \centering
    \includegraphics[width=0.75\textwidth]{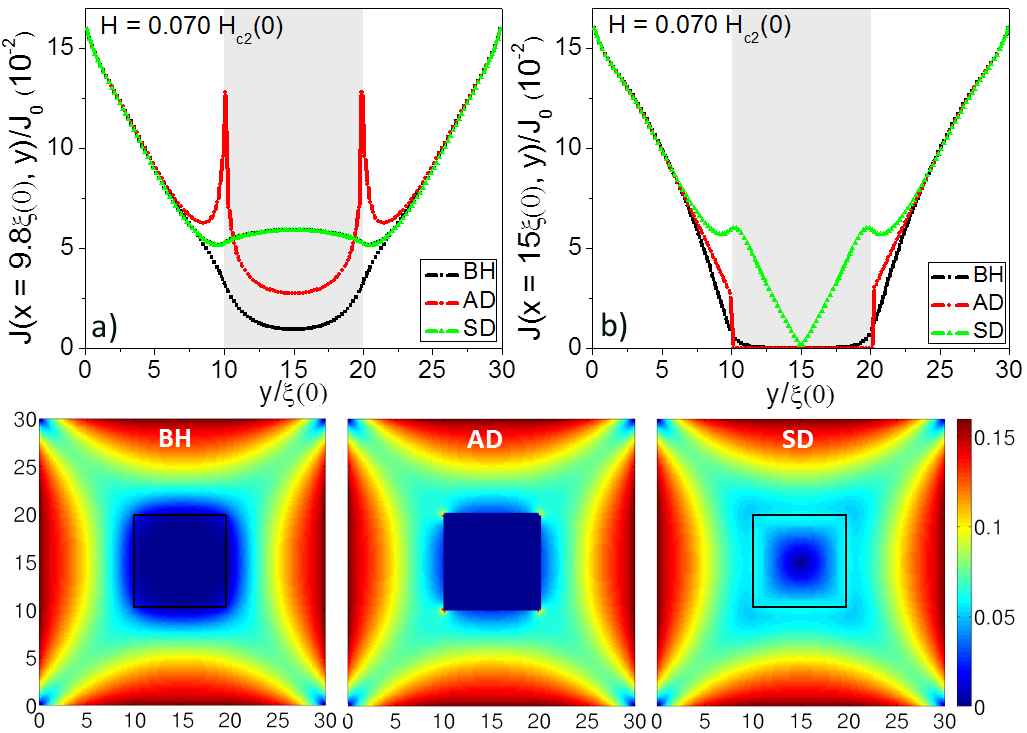}
    \caption{Superconducting current density as a function of the 
    position in the Meissner state for $H = 0.070 H_{c2} (0)$. 
    (a) Variation of $J$ along the $y$-axis for $x = 9.8 \xi(0)$ 
    (a line close to the edge of the defects). (b) Variation of $J$ 
    along the middle line $x = 15 \xi (0)$. Below these two panels, 
    it is shown the snapshots of $J$ for all the simulated samples. 
    The red dots on the vertices of the AD defect indicate the 
    current crowding effect at these points. For SD, $J$ 
    presents a local minimum in the defect's vicinity and a 
    linear dependence as a function of the position inside it.}
    \label{fig4}
\end{figure}

Such behavior could be understood by looking at the distribution of $J$ 
in  Meissner state, which is shown in Figure~\ref{fig4}, 
for $H = 0.070 H_{c2}(0)$. In panel (a) it is 
shown the current density distribution 
as a function of $y$ for $x = 9.8 \xi(0)$, i.e., along a 
line very close to the border of the defect. It can be 
seen that the current changes smoothly near the edge of 
the BH defect, whereas for the AD one, $J$ presents a 
dramatic variation near the corner, signaling a very 
strong current crowding~\cite{ClemBerggren2011}. In panel 
(b) of Figure~\ref{fig4} it is shown the variation of $J$ 
along a middle line $x = 15 \xi(0)$, going through the defects. 
As expected, while $J$ suddenly drops to zero inside the AD, 
it goes smoothly to zero at the BH. However, for SD, $J$ 
presents a local minimum outside the defect and a 
linear dependence on its position.

\begin{figure}[!h]
    \centering
    \includegraphics[width=0.75\textwidth]{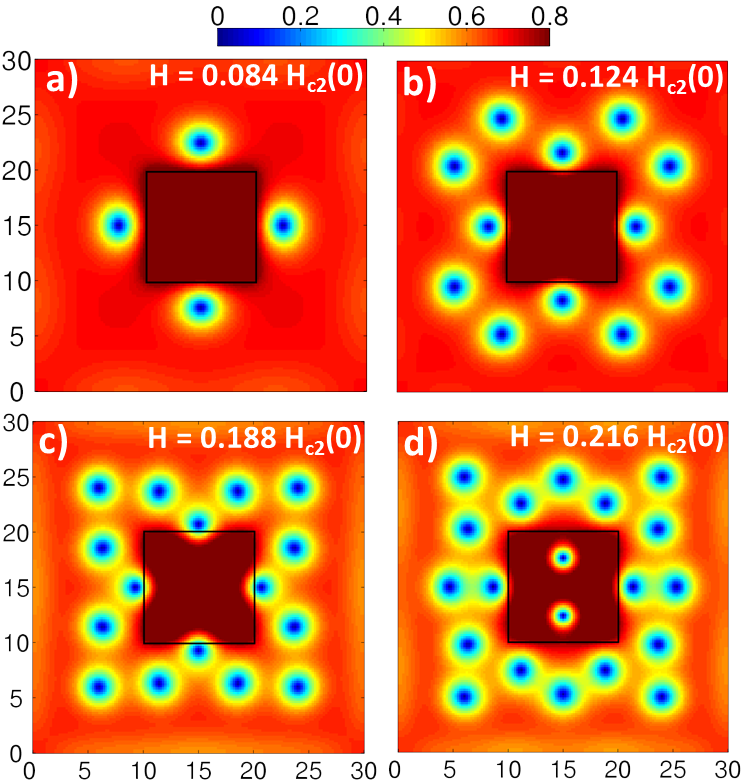}
    \caption{Superconducting order parameter $|\psi|$ in the 
    equilibrium state of the first four vortex penetrations 
    in the SD sample. (a) First vortex penetration in the 
    sample at $H = 0.084 H_{c2} (0)$; (b) second vortex 
    penetration at $H = 0.124 H_{c2} (0)$; (c) third vortex 
    penetration at $H = 0.188 H_{c2} (0)$; (d) fourth 
    vortex penetration in the sample at $H = 0.216 H_{c2} (0)$.}
    \label{fig5}
\end{figure}

The behavior of $J$ in the SD defect explains the late penetration 
of vortices inside it compared to the BH and AD defects. As such, 
defect presents better superconducting properties (higher $T_c$ and $J$) 
than the surrounding material, a substantial magnetic pressure must 
be applied to move the vortices in the matrix (which acts as a 
vortex reservoir) to the SD region. Figure~\ref{fig5} shows 
snapshots of $|\psi|$ for different values of $H$. Only for $H=0.216 H_{c2}(0)$ 
the first penetration of vortices will penetrate the SD defect.


\subsection{Non-equilibrium aspects - Characterization 
of the vortex dynamics}\label{subsec:subsec3.2}

\begin{figure}[!h]
    \centering
    \includegraphics[width=0.75\textwidth]{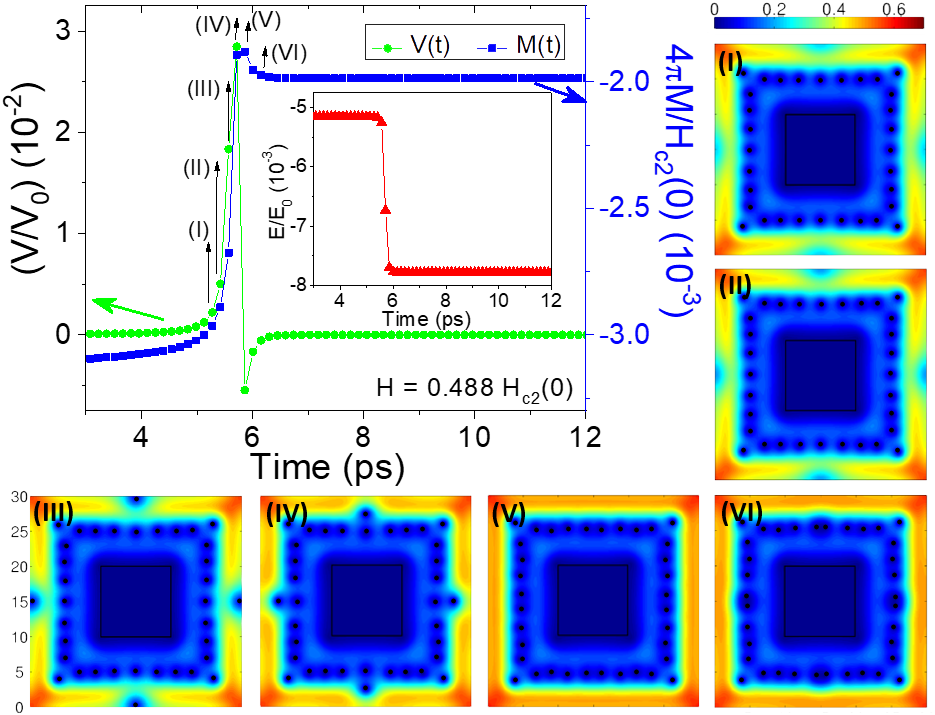}
    \caption{Time evolution of the induced voltage $V(t)$, magnetization $M(t)$, 
    and free energy $E(t)$ for the BH sample   at $H = 0.488H_{c2} (0)$. The 
    snapshots of $\psi$, enumerated from (I) to (VI),  show the configuration 
    of vortices along their dynamics.}
    \label{fig6}
\end{figure}

By the out of the equilibrium states, one 
can study the vortex dynamics by analyzing the temporal behavior 
of the induced voltage $V(t)$, magnetization $M(t)$, and free energy $E(t)$. 
The main panel of Figure~\ref{fig6} shows $V(t)$ and $M(t)$ 
for the vortex penetration at $H = 0.488H_{c2}(0)$ in BH sample. 
The inset shows the $E(t)$ behavior. The time, in picoseconds, 
was calculated taking into account a Pb-In alloy, with $T_c = 7~$K and $\kappa = 5$. 
The $V(t)$ curve is directly related to the vortex 
motion and accommodation. The voltage increases with the vortex nucleation, 
as can be seen in snapshots I, II, and III reaching a 
maximum value, $V = 2.85 \times 10^{-2} V_0$, with the 
complete vortex penetration ($t = 33~$ps). However, the 
strong repulsive interaction with the inner vortices 
(see snapshot IV) slows down the penetrating vortices 
leading $V(t)$ to a quick drop. After that, a small 
valley at $t = 34~$ps with a negative $V(t)$ ($V = -0.54 \times 10^{-2} V_0$) 
takes place. Such behavior is related to an accommodation 
of the vortices (exemplified by the snapshots V and VI in Figure~\ref{fig5}).

Still in Figure~\ref{fig6} is noticed a delay of $M(t)$ response 
comparing to the $V(t)$ one. The magnetization is also non-sensitive 
to the vortex accommodation. The filled squares curve in 
Figure~\ref{fig6} shows the $M(t)$ response where a peak is 
evidenced as a consequence of the vortex penetration. The inset at
the same figure shows the minimization of the free energy after 
the vortex penetration, and it is also non-sensitive to
vortex accommodations. The AD sample presents a similar 
behavior, and it is not shown here.

\begin{figure}[!h]
    \centering
    \includegraphics[width=0.75\textwidth]{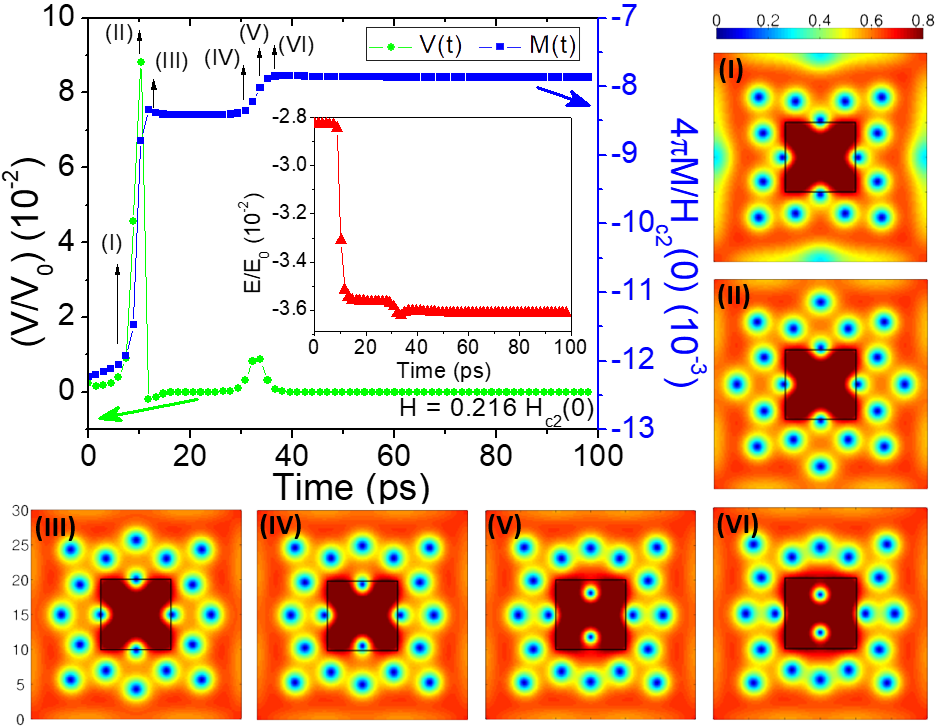}
    \caption{Time evolution of the induced voltage $V(t)$, 
    magnetization $M(t)$, and free energy $E(t)$ for the SD 
    sample at $H = 0.216H_{c2} (0)$. The snapshots of $\psi$, 
    enumerated from (I) to (VI), show the configuration of 
    vortices for some instants.}
    \label{fig7}
\end{figure}

The penetration dynamics at $H = 0.216H_{c2}(0)$ is shown in Figure~\ref{fig7} for the SD sample. It is noticed qualitative changes in $V(t)$, $M(t)$, and $E(t)$ responses in comparison with the results for BH and AD samples. In such a case, the $V(t)$ presents two peaks. The 
first at $t = 59~$ps ($V = 8.82 \times 10 ^{-2} V_0$) is associated with the penetration of four vortices as shown in the snapshot I and II. The second peak at $t = 195~$ps ($V = 0.87 \times 10^{-2} V_0$) is due to the penetration of two vortices in the defect, as shown by the sequence of snapshots from III to VI. The $M(t)$ increases with the penetration of the four other vortices, followed by a meta-stable equilibrium characterized by a plateau of about $100~$ps. After that, a transition to a steady-state occurs with vortex penetration into the defect. Besides that, those meta-stable and stable vortex configuration occurs due to a minimization of $E(t)$, which also shows two plateaus. 


\begin{figure}[!h]
    \centering
    \includegraphics[width=0.75\textwidth]{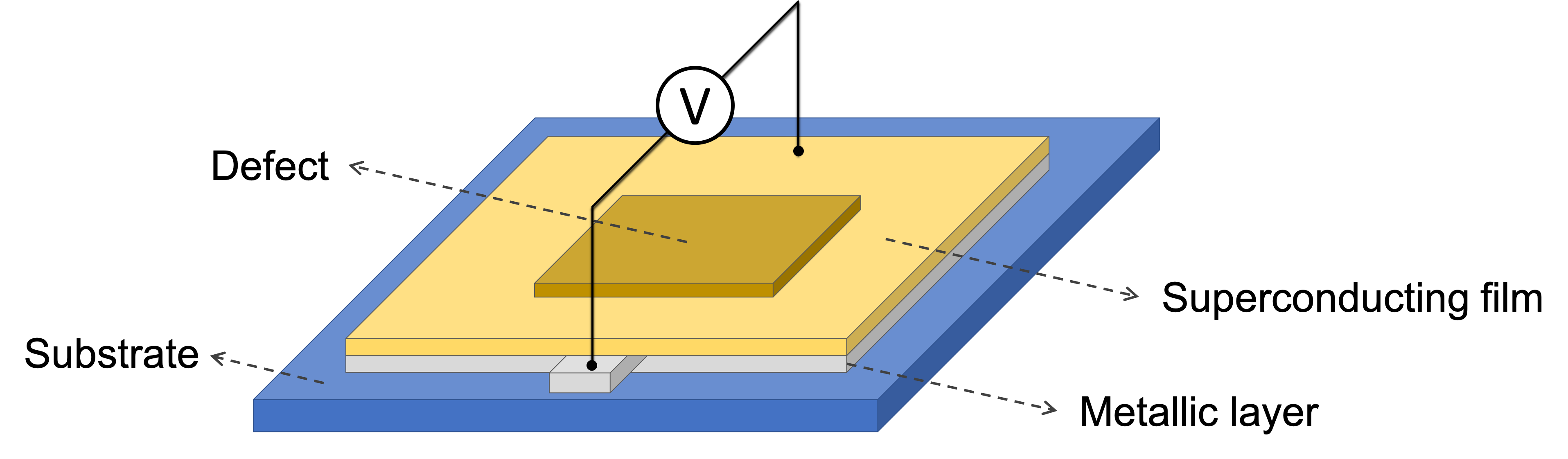}
    \caption{Suggestion of an experimental apparatus to monitor the 
    vortex motion in a decorated superconducting film. A metallic 
    layer coupled to a voltmeter is added to the system and the 
    voltage pulse as a consequence of vortex motion is captured.}
    \label{fig:coil}
\end{figure}

\textcolor{blue}{The time evolution of some physical properties of 
superconducting systems such as $V(t)$, $M(t)$, and $E(t)$ is a good 
tool to study the vortex dynamics. One possible way to capture 
experimentally the penetration and motion of vortices is preparing 
a superconducting film decorated with the corresponding defect by 
adding an adjacent metallic layer. Then, potential leads are 
connected to a voltmeter with time resolution and voltage scale 
to monitor the signal in this normal layer, as presented in 
Figure~\ref{fig:coil}. Once vortex motion occurs, the variation 
of magnetic flux induces eddy currents in the metal and a 
voltage pulse is measured. However, the time dependence of 
the induced electrical voltage in the metallic layer is not 
exactly the same as shown in Figures~\ref{fig6} and~\ref{fig7} 
due to the variation of the applied magnetic field. A similar 
superconductor-metal hybrid was prepared by P. Mikheenko and 
co-authors~\cite{Mikheenko2013} to obtain the velocity of 
flux avalanches, which is composed by a bunch of magnetic 
flux with extremely high velocities (as high as 100 km/s). 
Therefore, the voltage signal of some vortices is somehow 
far more challenging.}

\section{Conclusion}\label{sec:sec4}

We studied the effect of pinning centers of different natures on 
the vortex matter in mesoscopic superconductors. It was shown that 
$H_{c1}$ (the field for the first penetration) does not depend on 
the defect's nature. However, the following penetrations result 
from the competing currents distribution, i.e., the interaction 
between the shielding currents and the currents surrounding the 
defects. Such a fact also plays an essential role in the vortex 
arrangement in the superconducting region. Also, the crowding of 
the currents produces a less degraded superconducting region 
around the antidot, contrasting with a blind hole. 
Such a remaining superconducting region could drive the 
sample with an antidot to support a higher $H_{c2}$ than 
samples with blind holes. Besides, more vortices populate 
the sample with a defect made of a superconductor with 
higher $T_c$ (strong defect) at the first penetrations. 
This defect could protect the inner material and control 
the vortex dynamics with more accuracy in that region. 
It was also shown that $M(t)$ is not sensitive to vortex 
accommodation but indicates the penetrations in the 
strong defect. On the other hand, $V(t)$ is sensitive 
to vortex penetration and accommodation, an alternative 
tool for studying vortex dynamics in mesoscopic superconducting 
systems.

\ack{
We thank the Brazilian Agencies S\~ao Paulo Research 
Foundation (FAPESP), grants 2016/12390-6, 2018/06575-9, and 2020/10058-0, 
Coordena\c c\~ao de Aperfei\c coamento de Pessoal de N\'ivel 
Superior - Brasil (CAPES) - Finance Code 001, National Council 
of Scientific and Technological Development (CNPq, grant 302564/2018-7).}

\section*{References}
\bibliographystyle{iopart-num}
\bibliography{references}

\end{document}